\documentclass[aps,prl,reprint,twocolumn,superscriptaddress,english]{revtex4}
\usepackage{amssymb}
\usepackage{graphicx}
\usepackage{siunitx}
\usepackage{amsmath}
\usepackage{amsthm}
\usepackage{amsfonts}
\usepackage[T1]{fontenc}
\usepackage[latin9]{inputenc}
\usepackage{array}
\usepackage{multirow}
\usepackage{color}
\usepackage{esint}
\usepackage{bm}
\usepackage{bbm}
\usepackage{hyperref}
\usepackage{babel}
\usepackage{titlesec}
\hypersetup{pdfpagemode=FullScreen,colorlinks=true,breaklinks,urlcolor=blue,linkcolor=blue,citecolor=blue}

\begin{document}

\title{Observing topological charges and dynamical bulk-surface correspondence with ultracold atoms}

\author{Chang-Rui Yi}
\affiliation{Hefei National Laboratory for Physical Sciences at Microscale
and Department of Modern Physics, University of Science and Technology of China, Hefei, Anhui 230026, China}
\affiliation{Shanghai Branch, CAS Center for Excellence and Synergetic Innovation Center in Quantum Information and Quantum Physics, University of Science and Technology of China, Shanghai 201315, China}

\author{Long Zhang}
\author{Lin Zhang}
\affiliation{International Center for Quantum Materials, School of Physics, Peking University, Beijing 100871, China}
\affiliation{Collaborative Innovation Center of Quantum Matter, Beijing 100871, China}
\author{Rui-Heng Jiao}
\author{Xiang-Can Cheng}
\author{Zong-Yao Wang}
\author{Xiao-Tian Xu}
\author{Wei Sun}

\affiliation{Hefei National Laboratory for Physical Sciences at Microscale
and Department of Modern Physics, University of Science and Technology of China, Hefei, Anhui 230026, China}
\affiliation{Shanghai Branch, CAS Center for Excellence and Synergetic Innovation Center in Quantum Information and Quantum Physics, University of Science and Technology of China, Shanghai 201315, China}

\author{Xiong-Jun Liu}
\email{xiongjunliu@pku.edu.cn}
\affiliation{International Center for Quantum Materials, School of Physics, Peking University, Beijing 100871, China}
\affiliation{Collaborative Innovation Center of Quantum Matter, Beijing 100871, China}

\author{Shuai Chen}
\email{shuai@ustc.edu.cn}
\author{Jian-Wei Pan}
\email{pan@ustc.edu.cn}
\affiliation{Hefei National Laboratory for Physical Sciences at Microscale
and Department of Modern Physics, University of Science and Technology of China, Hefei, Anhui 230026, China}
\affiliation{Shanghai Branch, CAS Center for Excellence and Synergetic Innovation Center in Quantum Information and Quantum Physics, University of Science and Technology of China, Shanghai 201315, China}


\date{\today}

\begin{abstract}
In quenching a topological phase across phase transition, the dynamical bulk-surface correspondence emerges that the bulk topology of $d$-dimensional ($d$D) phase relates to the nontrivial pattern of quench dynamics emerging on $(d-1)$D subspace, called band inversion surfaces (BISs) in momentum space. Here we report the first experimental observation of the dynamical bulk-surface correspondence through measuring the topological charges in a 2D quantum anomalous Hall model realized in an optical Raman lattice.
The system can be quenched with respect to every spin axis by suddenly varying the two-photon detuning or phases of the Raman couplings, in which
the topological charges and BISs are measured dynamically by the time-averaged spin textures.
We observe that the total charges in the region enclosed by BISs define a dynamical topological invariant,
which equals the Chern index of the post-quench band. The topological charges relate to an emergent dynamical field which exhibits nontrivial topology on BIS, rendering the dynamical bulk-surface correspondence.
This study opens a new avenue to explore topological phases dynamically.
\end{abstract}

\maketitle

{\em Introduction.---}
Topological quantum matter~\cite{topo_insu,insu_conductor} has attracted intense interest due to the discovery of new fundamental phases~\cite{QSHIS,ExpQAHE,ExpWeyl} and broad potential applications~\cite{topoFlatBand,quantum_compu}.
Recent experimental advances in cold atoms highlight the realizations of various topological models,
such as the one-dimensional (1D) Su-Schrieffer-Heeger model~\cite{ZakPhase},
1D chiral topological phase~\cite{topoPhase_1D}, and 2D Chern insulator~\cite{HofHamBloch,HofHamKett,HaldaneModel,HofBandC,realization2DSOC,W.S_realization2DSOC}.
The studies commonly faces an important question: how to measure topological indices for cold atom systems?
The 1D winding number can be detected by measuring Zak phase via Ramsey interferometry~\cite{ZakPhase}.
In 2D spin-orbit (SO) coupled Chern phase~\cite{realization2DSOC,W.S_realization2DSOC},
the Chern number can be determined by measuring Bloch states at highly symmetric momenta~\cite{Liu2013PRL-b}.
These methods are however not generic or lack sufficient accuracy.

Recently, a research focus has been drawn to non-equilibrium dynamics in topological quantum phases
~\cite{topoPhase_1D,quenchVortice,BerryCurvature,PhysRevLett.119.080501,PhysRevLett.117.126803,PhysRevLett.117.235302}. Several theoretical works~\cite{linking_HuiZhai,windingNumber_zhang,Hopflinks,topoCharge,zhang2019emergent} proposed dynamical characterizations of topological phases by quantum quenches, with
some predictions having been studied in experiment~\cite{uncover_topology,link2019,HopfFibration}.
In particular, a dynamical bulk-surface correspondence was proposed~\cite{windingNumber_zhang}, showing that the bulk topology of a $d$D topological phase universally corresponds to the nontrivial pattern of quench dynamics emerging on the $(d-1)$D momentum subspace
called band inversion surfaces (BISs), analogy to the well-known bulk-boundary correspondence in real space~\cite{topo_insu,insu_conductor}.
A recent experiment observed the ring pattern of BISs dynamically~\cite{uncover_topology}. However, the topological invariant of quench dynamics emerging on BISs was not observed, so the experimental verification of the essential dynamical bulk-surface correspondence is yet to be pursued.

In this letter, we report the experimental observation in ultracold atoms of the dynamical bulk-surface correspondence~\cite{windingNumber_zhang} following a new scheme proposed in Ref.~\cite{topoCharge}, and characterize the topological phases by dynamically detecting topological charges of monopoles in momentum space. The central idea of the new scheme is that through a sequence of quantum quenches along all spin axes in the topological system, the topology can be detected by measuring the quantum dynamics for only $z$-component spin polarization in each quench. We implement the study in a 2D quantum anomalous Hall (QAH) model in an optical Raman lattice, and quench the system along different spin axes by quickly varying the two-photon detuning or phases of Raman couplings based on the new scheme~\cite{topoCharge}.
The complete information of bulk topology, including the topological charges and BISs, is extracted by measuring the $z$-component spin dynamics.
We observe that the total charges in the region enclosed by BISs equal the Chern number of the post-quench phase, and are related to an emergent dynamical field which exhibits nontrivial topological pattern on the BIS, rendering the dynamical bulk-surface correspondence in momentum space.

\begin{figure}
\includegraphics[width=0.95\linewidth]{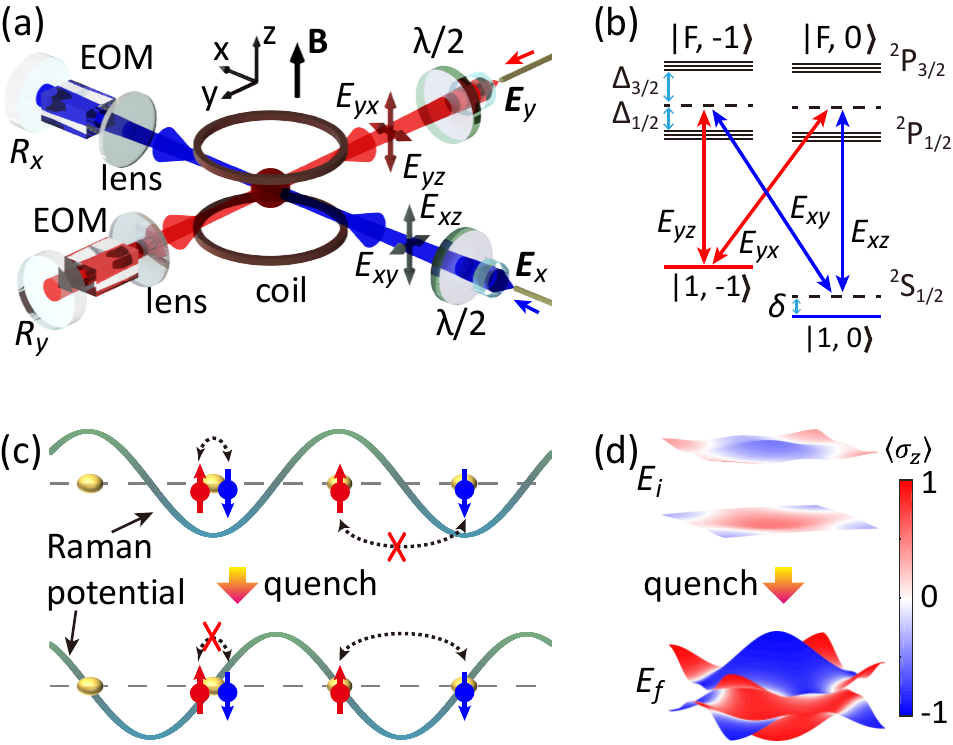}
\caption{
Experimental scheme.
(a) Experimental setup. Two laser beams $\bm{E}_x$ and $\bm{E}_y$ are incident on the atoms and then reflected by two mirrors $R_{x,y}$,
simultaneously producing a 2D square lattice and two Raman coupling potentials. The $\lambda/2$ wave plates are used to generate
two orthogonally polarized components $\bm{E}_{xy,xz}$ or $\bm{E}_{yx,yz}$  for each beam, and
EOMs are placed in front of the mirrors to tune the phase shift between the
two components.
(b) Level structure and Raman transitions.
Two hyperfine states $\left|1, -1\right\rangle$  and $\left| 1, 0\right\rangle$ are selected to form the double-$\Lambda$-type coupling configuration.
(c) The quench process corresponds to changing the symmetry of Raman potentials (sinusoidal curves) with respect to
lattice sites (yellow ellipsoids) in a spatial direction. The relative (anti)symmetry selects the hopping type:
Before quench, only on-site spin flipping is permitted while spin-flipped hopping is forbidden; after quench the situation is reversed.
(d) The $s$-band structure before and after quenching $h_{x}$ or $h_{y}$.
The color indicates the distribution of spin polarization $\langle\sigma_z\rangle$ with respect to eigenstates of the pre- or post-quench
Hamiltonian.
}
\label{fig1}
\end{figure}

{\em Experimental setup.---}
Our experiment is based on a 2D QAH model~\cite{Liu2014PRL-a} with $C_4$ symmetry proposed in theory~\cite{WangBaoZong2018} and realized for ultracold $^{87}\textrm{Rb}$ bosons trapped in a square optical Raman lattice~\cite{realization2DSOC,map2DSOCband,W.S_realization2DSOC}.
The laser beam ${\bm E}_x$ (${\bm E}_y$) with wavelength of
$\lambda={787}nm$ is incident from $x$ ($y$) direction and reflected by mirror.
The two beams pass through $\lambda/2$ wave plates and each splits into two orthogonally
polarized components ${\bm E}_x=\bm{E}_{xz}+\bm{E}_{xy}$ and ${\bm E}_y=\bm{E}_{yz}+\bm{E}_{yx}$ [Fig.~\ref{fig1}(a)], with ${\bm E}_{xy}=\hat y\left|E_{xy}\right|\cos(k_0x)$,
${\bm E}_{xz}=\hat z\left|E_{xz}\right|e^{i\varphi_{1}}\cos(k_0x-\varphi_{1})$,
${\bm E}_{yx}=\hat x\left|E_{yx}\right|\cos(k_0y)$, and
${\bm E}_{yz}=\hat z\left|E_{yz}\right|e^{i\varphi_{2}}\cos(k_0y-\varphi_{2})$,
from which the scalar and vector potentials are generated and give the lattice and Raman potentials, respectively~\cite{W.S_realization2DSOC,WangBaoZong2018}.
Here $k_0=2\pi/\lambda$ and $\varphi_{1}$ ($\varphi_{2}$) is the relative phase between $\bm{E}_{xz}$ ($\bm{E}_{yz}$) and $\bm{E}_{xy}$ ($\bm{E}_{yx})$. As proposed in Ref.~\cite{topoCharge}, the key technique here is that we apply electro-optic modulators (EOMs) to manipulate the relative symmetries between lattice and Raman potentials by tuning $\varphi_{1,2}$.
The spin-$1/2$ system, selected from the hyperfine states $\left| \uparrow \right\rangle=\left| F=1, m_F=-1\right\rangle$ and $\left| \downarrow \right\rangle=\left|1, 0\right\rangle$, are coupled by the Raman transitions
[see Fig.~\ref{fig1}(b)]. In tight-binding regime the realized Bloch Hamiltonian ${\cal H}({\bm q})={\bm h}({\bm q})\cdot{\bm\sigma}$ reads (see more details in supplementary materials~\cite{supplementary})
\begin{align}\label{Hq}
{\cal H}({\bm q})=&(m_x+2t^{y}_{\rm so}\sin q_y)\sigma_x+(m_y+2t^x_{\rm so}\sin q_x)\sigma_y\nonumber\\
&+(m_z-2t^x_0\cos q_x-2t^y_0\cos q_y)\sigma_z,
\end{align}
where ${\bm q}=(q_x,q_y), m_i,\sigma_i$, and $t^{i}_{0}$ ($t_{\rm so}^{i}$) are respectively the Bloch momentum, Zeeman constants,
Pauli matrices acting on spin space, and the spin-conserved (flip) hopping coefficients, with $i=x,y,z$. The $m_z$-term is related to the two-photon detuning by $m_z=\delta/2$.
The Zeeman constants $m_{x,y}$ and spin-flip hopping coefficients $t_{\rm so}^{x,y}$ are induced by the Raman couplings and controlled by $\varphi_{1,2}$.
In this experiment, the quenches along different spin axes will be performed by quickly modulating $m_{x,y,z}$~\cite{supplementary,topoCharge}.

The two Raman potentials are generated by two pairs of beams ($\bm{E}_{xz},\bm{E}_{yx}$) and ($\bm{E}_{yz}, \bm{E}_{xy}$), respectively.
For the setting with $(\varphi_1,\varphi_2)=(\pi/2,\pi/2)$, the former (or latter) potential $\propto\sin(k_0x)\cos(k_0y)$ (or $\sin(k_0y)\cos(k_0x)$) is symmetric with respect to every lattice site
in $y$ (or $x$) direction, but antisymmetric in $x$ (or $y$) direction~\cite{supplementary}. This is the configuration of 2D optical Raman lattice to realize QAH model~\cite{WangBaoZong2018,realization2DSOC,W.S_realization2DSOC,map2DSOCband}, in which the nearest-neighbor spin-flip hopping is induced by each Raman potential along the antisymmetric direction, while on-site spin-flip transition is forbidden. We then reach the Hamiltonian~\eqref{Hq} with $m_x=m_y=0$. Further,
for the other setting with $(\varphi_1,\varphi_2)=(0,\pi/2)$ (or $(\pi/2,0)$), the former (or latter) Raman potential $\propto\cos(k_0x)\cos(k_0y)$, which is symmetric with respect to each lattice site in both $x$ and $y$ directions, and induces on-site spin-flip transition, but no nearest-neighbor spin-flip hopping. Then a nonzero $m_y$ (or $m_x$) is induced, while $t^{x}_{\rm so}\approx0$ (or $t^{y}_{\rm so}\approx0$)~\cite{topoCharge,supplementary}, and
the Hamiltonian (\ref{Hq}) can describe a trivial system with large transverse Zeeman field.

\begin{figure}
\includegraphics[width=\linewidth]{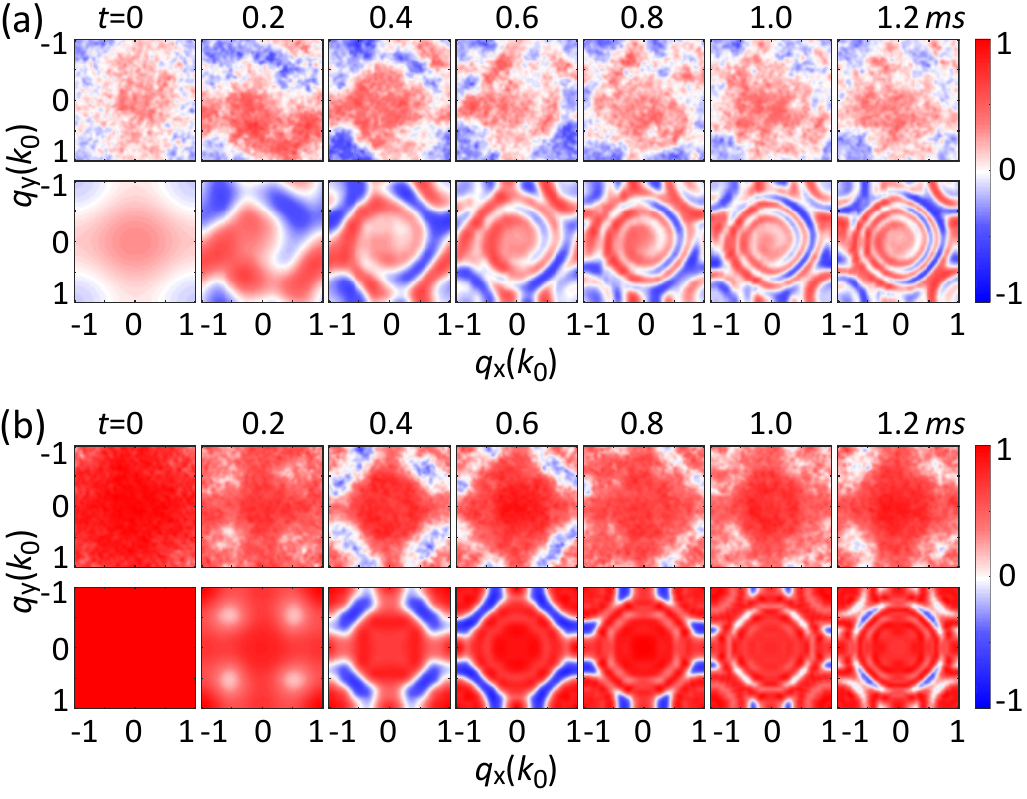}
\caption{Time evolution of the spin polarization $\langle\sigma_z({\bm q})\rangle_{y,z}$ after quenching $h_{y}$ (a) and $h_{z}$ (b), respectively.
Quenching $h_y$ is realized by tuning the phases from $(\varphi_1,\varphi_2)=(0,\pi/2)$ to the setting $(\pi/2,\pi/2)$.
Quenching $h_z$ corresponds to varying the two-photon detuning from $\delta=-200E_{\rm r}$ to $-0.2E_{\rm r}$.
In each case, the measured spin textures for different hold time $t$ (upper) are compared with numerical calculations (lower).
Spin textures after the two quenches exhibit distinctive patterns during the time evolution.
The post-quench parameters are fixed at lattice depth $V_0=4.0E_{\rm r}$, Raman coupling strength $\Omega_0=1.0E_{\rm r}$ and
$\delta=-0.2E_{\rm r}$.
Here $E_{\rm r}$ is the recoil energy.
}
\label{fig2}
\end{figure}

\begin{figure*}
\includegraphics[width=\linewidth]{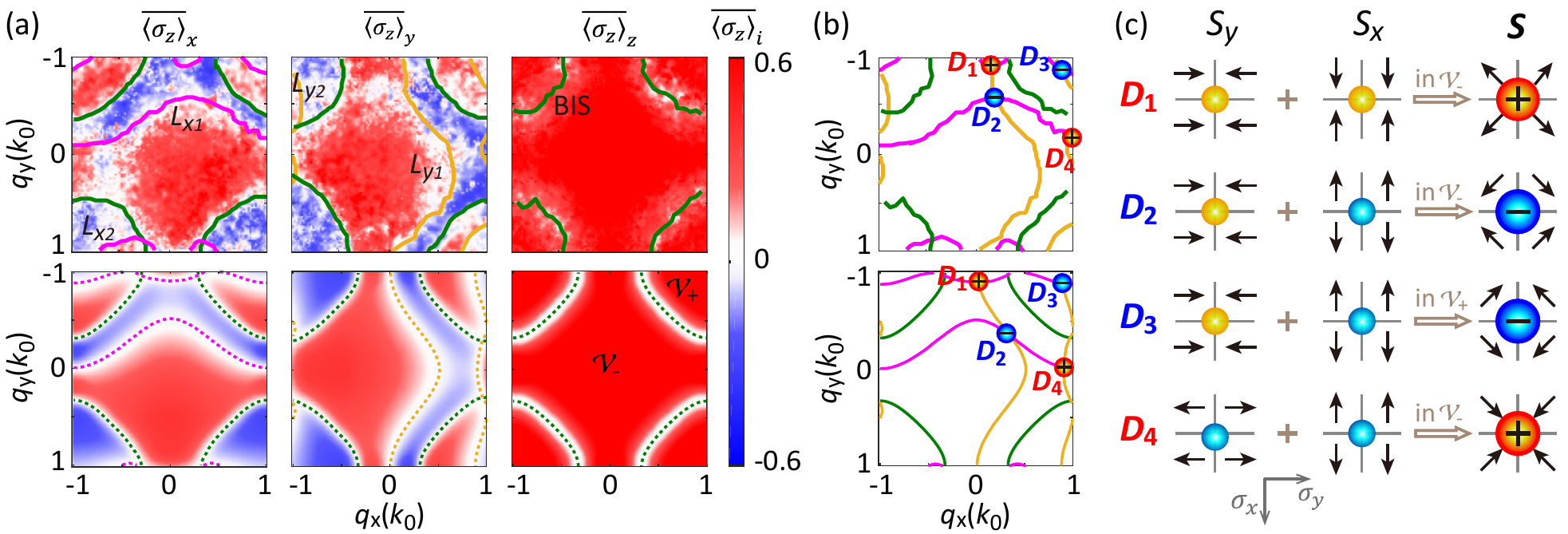}
\caption{
Dynamical measurement of topological charges and Chern number.
(a) Experimental measurements (upper) of time-averaged spin textures $\overline{\left \langle \sigma_{z}(\bm{q}) \right \rangle}_{i}$
by quenching $h_i$ ($i=x,y,z$), compared with numerical calculations (lower). Dynamical characterization of
$\overline{\langle\sigma_{z}({\bm q})\rangle}_z=0$ identifies the BIS (green), which also emerges in
$\overline{\langle\sigma_{z}({\bm q})\rangle}_{x,y}$. The BIS divides the BZ into two regions: ${\cal V}_{-}$ with $h_z<0$ and ${\cal V}_{+}$ with $h_z>0$.
 In both $\overline{\langle\sigma_{z}({\bm q})\rangle}_{x}$ and $\overline{\langle\sigma_{z}({\bm q})\rangle}_{y}$, other than the BIS,
two additional curves are formed at momenta with vanishing spin polarization, denoted by $L_{x1,x2}$ (magenta) and $L_{y1,y2}$ (brown), respectively. (b) The four lines
$L_{x1,x2}$and $L_{y1,y2}$ have four intersections $D_{1,2,3,4}$, marking the locations of topological charges.
Three charges are in the region ${\cal V}_{-}$ and one in ${\cal V}_{+}$.
(c) The charge value ${\cal C}_i=\pm1$ at each intersection $D_i$ ($i=1,2,3,4$) is determined by the signs of spin textures
$\overline{\langle\sigma_{z}({\bm q})\rangle}_{x,y}$ in the neighboring four subregions. The plus (minus) sign of
$\overline{\langle\sigma_{z}({\bm q})\rangle}_{x,y}$ indicates that the
field components $S_{x,y}$ point to the positive (negative) spin axes.
When ${\bm S}$ flows from or to the charge, the charge value ${\cal C}_i=+1$ (red); if ${\bm S}$ is ``repulsed'' by the charge, ${\cal C}_i=-1$ (blue). Three topological charges are enclosed in ${\cal V}_{-}$ with total charges being ${\cal C}=+1$, giving the Chern number ${\rm Ch}=+1$.
The post-quench parameters are $V_0=4.0E_{\rm r}, \Omega_0=1.0E_{\rm r}$ and $\delta=-0.2E_{\rm r}$.
}\label{fig3}
\end{figure*}

{\em Quenches and measurements.---}
We implement the quenches with respect to all spin quantization axes. For simplicity we consider that $t^{x}_{\rm so}=t^{y}_{\rm so}=t_{\rm so}$ and $t^{x}_0=t^{y}_0=t_0$ by setting the lattice depths ($V_0$) and the
Raman potential amplitudes ($\Omega_0$) to be isotropic for post-quench Hamiltonian.
Quenching $h_z$ is realized by tuning the two-photon detuning $\delta$~\cite{uncover_topology,supplementary}, while
quenching $h_{x,y}$ is performed by suddenly tuning $(\varphi_1,\varphi_2)$ between the aforementioned two settings [see Fig.~\ref{fig1}(c)].
In each quench, the post-quench parameters are fixed.
Quenching $h_{x,y}$ is performed as follows.
First, the $^{87}\textrm{Rb}$ atoms are prepared slightly above the critical temperature of Bose-Einstein condensation, and
then adiabatically loaded into the Raman lattice with ($\varphi_1,\varphi_2)=(0,\pi/2)$ or $(\pi/2,0)$ by ramping up
the laser intensity in ${100}ms$~\cite{supplementary}. The atoms are then populated almost in the lowest trivial energy band of the pre-quench Hamiltonian [Fig.~\ref{fig1}(d)].
Second, using the EOM in the ${x}$ (or ${y}$) direction we suddenly tune the phase to shift as
($\varphi_1,\varphi_2)\to(\pi/2,\pi/2$) within ${2}\mu s$,
corresponding to quenching $h_y$ (or $h_x$).
The atoms are driven out of equilibrium and evolve under the post-quench Hamiltonian with the topologically nontrivial bands [Fig.~\ref{fig1}(d)].
Third, we hold the system for a certain time $t$, and measure the atom density $n_{\uparrow,\downarrow}({\bm{q}},t)$ of each spin component in momentum space by time-of-flight (TOF) expansion. The evolution of spin polarization $\langle\sigma_z({\bm q},t)\rangle=[n_{\uparrow}({\bm{q}},t)-n_{\downarrow}({\bm{q}},t)]/[n_{\uparrow}({\bm{q}},t)+n_{\downarrow}({\bm{q}},t)]$ is obtained at each Bloch momentum.

Figure~\ref{fig2} shows the measured spin textures $\langle\sigma_z({\bm q},t)\rangle_{y,z}$ at different time after quenching $h_{y,z}$, which are compared with the numerical calculations. The spin evolution $\langle\sigma_z({\bm q},t)\rangle_{x}$ is also measured by quenching $h_x$ but not shown here for simplicity.
The polarizations $\langle\sigma_z({\bm q},t)\rangle_{y,z}$ oscillate at frequencies equaling to local gap of the post-quench bands at momentum $\bm{q}$. In particular, for quenching $h_y$ in (a), there appears a spiral-like dynamical pattern, which gradually spreads to the whole BZ.
For quenching $h_z$ in (b), the spin oscillations exhibit a ring structure with $C_4$ symmetry, which
is the dynamical signature of the BIS emerging in quench dynamics.
For each quench, we fit the spin oscillation at every momentum ${\bm q}$~\cite{supplementary}, with which we further obtain the
time-averaged spin textures $\overline{\left \langle \sigma_{z}(\bm{q}) \right \rangle}_{x,y,z}$.
These spin textures exhibit nontrivial patterns which characterize the topology of post-quench band, as elaborated below.

\emph{Measuring topological charges.---}
As shown in theory~\cite{windingNumber_zhang,topoCharge}, the bulk topology can be characterized by the total
topological charges in the region enclosed by BISs.
The information of topological charges and BISs is captured dynamically by the time-averaged spin textures $\overline{\left \langle \sigma_{z}(\bm{q}) \right \rangle}_{i}$, which are shown in Fig.\ref{fig3}(a).
The momenta where all the polarizations $\overline{\left \langle \sigma_{z}(\bm{q}) \right \rangle}_{x,y,z}$ being zero form the BIS, which
shapes a ring around the $M$ point  ($q_x=q_y=k_0$), and divide the whole BZ into ${\cal V}_+$ (with $h_z>0$) and ${\cal V}_-$ (with $h_z<0$) regions~\cite{supplementary}.
Besides the BIS, there are also other momenta satisfying $\overline{\langle\sigma_{z}(\bm{q})\rangle}_{x}=0$ or $\overline{\langle\sigma_{z}(\bm{q})\rangle}_{y}=0$, and
form the curves denoted as $L_{x1,x2}$ and $L_{y1,y2}$, respectively. As shown in Fig.\ref{fig3}(b), the curves $L_{x1,x2}$ and $L_{y1,y2}$
intersect at four momentum points $D_{1,2,3,4}$ in the first BZ, which mark the locations of dynamical topological charges~\cite{topoCharge,TopoQuenchGeneral}: three in ${\cal V}_-$ and one in ${\cal V}_+$.
All our experimental measurements agree well with numerical calculations (Fig.\ref{fig3}), which are based on the full-band model and take into account the finite temperature effect~\cite{supplementary}. A special case is that if the pre-quench trivial phase is fully polarized by a large initial Zeeman term with $|m_i|\gg|t_{0,\rm so}|$, the BIS refers to the 1D band-crossing ring with $h_z=0$, and topological charges are located at the nodes of SO coupling with $h_x=h_y=0$, which are $\Gamma, X_{1,2}$ and $M$ points in the BZ~\cite{topoCharge}.

The charge value ${\cal C}=\pm1$ can be further characterized by the winding of the spin-texture field ${\bm S}({\bm q})=(S_y,S_x)$,
with components given by~\cite{topoCharge}
\begin{align}
S_{y,x}(\bm{q})=\frac{1}{{\cal N}_q}{\rm sign}[h_z({\bm q})]\overline{\langle\sigma_{z}(\bm{q})\rangle}_{y,x}.
\end{align}
Here ${\cal N}_q$ denotes the normalization factor.
Since the two curves that intersect at a charge divide the neighboring region into four parts, the charge value ${\cal C}$
can be simply determined by the signs of spin textures $\overline{\langle\sigma_{z}(\bm{q})\rangle}_{x,y}$ in the four subregions, as illustrated in Fig.~\ref{fig3}(c).
The plus (minus) sign of $\overline{\langle\sigma_{z}({\bm q})\rangle}_{x,y}$ indicates that the field components $S_{x,y}$ point to the positive (negative) spin axes.
The combined dynamical field ${\bm S}({\bm q})$ characterizes the charge ${\cal C}=+1$ if the field flows from or to the charge;
${\cal C}=-1$ if it is ``repulsed'' by the charge.
We finally find that three dynamical topological charges are enclosed in the region ${\cal V}_-$ with one being ${\cal C}=-1$ and two ${\cal C}=+1$,
whose summation characterizes the post-quench topological phase with the Chern number ${\rm Ch}=+1$.

\emph{Dynamical bulk-surface correspondence.---}
Similar to the well-known bulk-boundary correspondence in real space,
there is also a dynamical bulk-surface correspondence emerging from quench dynamics in the momentum space~\cite{windingNumber_zhang}.
It claims that the bulk topology can be characterized by the dynamical topological pattern on the lower-dimensional BISs. In Ref.~\cite{uncover_topology},
we have observed the quench dynamics that occurs on BISs, and accordingly determined the topological phase diagram,
but the topological pattern can not be measured by quenching only $h_z$.
Here, using the time-averaged spin textures obtained by a sequence of quenches,
we can observe the winding of an emergent dynamical field defined on the BIS, as an experimental observation of
the dynamical bulk-surface correspondence.

\begin{figure}
\includegraphics[width=\linewidth]{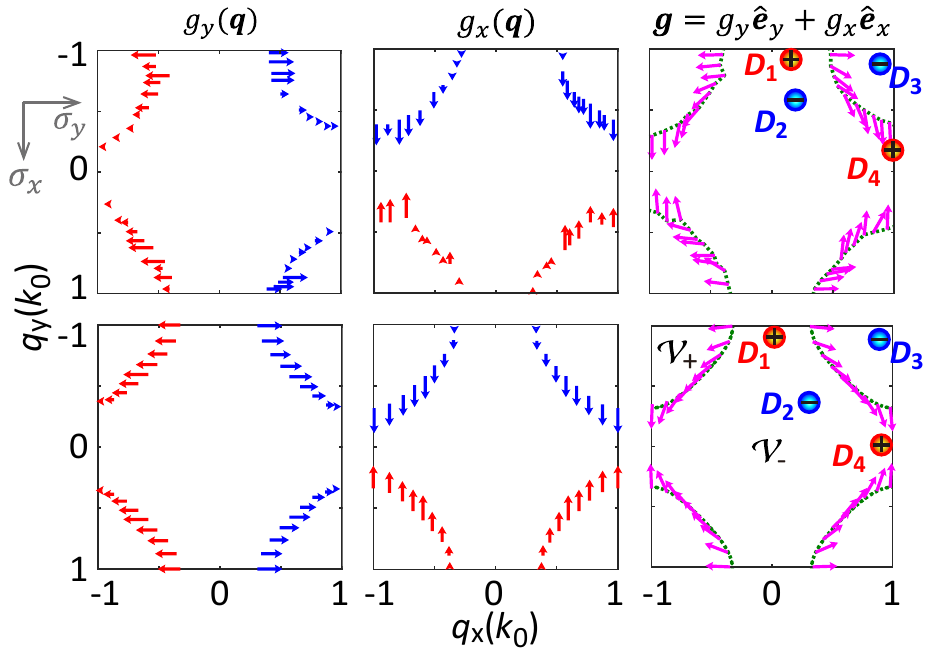}
\caption{Dynamical topological pattern emerging on the BIS.
The field component $g_i(\bm{q})$ ($i=x,y$) is obtained by the variation of time-averaged spin texture
$\overline{\langle\sigma_{z}(\bm{q})\rangle}_i$ [shown in Fig.~\ref{fig3}(a)] across the BIS.
The first (second) row are the experimental (numerical) results.
The measured dynamical field ${\bm g}(\bm{q})$ exhibits a nonzero winding along the ring, which
characterizes the bulk topology with Chern number ${\rm Ch}=+1$.
}
\label{fig4}
\end{figure}

According to Refs.~\cite{windingNumber_zhang,topoCharge},
the dynamical field $\bm{g}(\bm{q})=(g_y,g_x)$ can be defined by
\begin{align}\label{gxy}
g_{y,x}(\bm{q})\equiv\frac{1}{{\cal N}_{\bm q}}\partial_{q_{\perp}}\overline{\langle\sigma_{z}(\bm{q})\rangle}_{y,x},
\end{align}
where $\mathcal{N}_{\bm q}$ is normalization factor,
and $q_{\perp}$ denotes the momentum perpendicular to the BIS and points from ${\cal V}_{-}$ to ${\cal V}_{+}$.
We calculate the field components $g_{i}(\bm{q})$ ($i=x,y$) by the variations of spin textures
$\overline{\langle\sigma_{z}(\bm{q})\rangle}_{i}$ along the direction $q_{\perp}$, i.e.,
$g_{i}(\bm{q}_0)=\lim_{q_{\perp}\to0}[\overline{\langle\sigma_{z}(\bm{q}_0+q_{\perp}\hat{e}_{\perp})\rangle}_{i}-\overline{\langle\sigma_{z}(\bm{q}_0-q_{\perp}\hat{e}_{\perp})\rangle}_{i}]$
for every $\bm{q}_0$ on the BIS. The results are shown in Fig.~\ref{fig4}.
A combination of the two components gives the emergent dynamical field $\bm{g}(\bm{q})$, which is
observed to exhibit a nonzero winding along the ring despite local fluctuations induced by experimental errors.
The winding number of the field $\bm{g}(\bm{q})$, resembling the flux of magnetic monopoles,
counts the total charges ${\cal C}=+1$ in the region ${\cal V}_-$, and characterizes the post-quench
quantum phase with Chern number ${\rm Ch}=+1$.

\emph{Conclusion.---}
We have detected by quench dynamics the topological charges and further the dynamical bulk-surface correspondence in a 2D QAH model through a sequence of quantum quenches.
We observed that the bulk topology can be generally characterized by the total charges enclosed by a lower-dimensional momentum
subspace called band inversion surface (BIS).
Such topological charges are related to the emergent topological pattern of a dynamical field defined on the BIS, showing
an essential dynamical bulk-surface correspondence.
We emphasize that the scheme of quenching all the spin axes enables
to extract the complete topological information by
measuring only a single spin component, hence has great advantages in detecting
topological states.
This study clearly verifies the insight that a higher-dimensional topological
system (e.g. 2D Chern insulator) can be characterized by
lower-dimensional dynamical invariants (1D ring topology or 0D monopole charges),
which has broad applications to dynamical classification of generic topological
phases~\cite{windingNumber_zhang,topoCharge}.

{\it Note added.}--In completing the manuscript, we notice a preprint which demonstrates the dynamical bulk-surface correspondence based on nitrogen-vacancy center~\cite{wang2019experimental}. In comparison, we first measured the topological charges with ultracold atoms, and provide an alternative powerful approach to verify the dynamical bulk-surface correspondence.

\emph{Acknowledgement.---}
We thank Jin-Long Yu for fruitful discussions.
This work was supported by the National Key R\&D Program of China (under grants 2016YFA0301601 and 2016YFA0301604), National Natural Science Foundation of China (under grants No. 11674301, 11574008, 11825401, and 11761161003), the Chinese Academy of Sciences and the Anhui Initiative in Quantum Information Technologies (AHY120000) and the Chinese Academy of Sciences, and the Strategic Priority Research Program of Chinese Academy of Science (Grant No. XDB28000000).

Chang-Rui Yi  and Long Zhang contribute equally to this work.

%

%
%
\newpage
\onecolumngrid
\renewcommand\thefigure{S\arabic{figure}}
\setcounter{figure}{0}
\renewcommand\theequation{S\arabic{equation}}
\setcounter{equation}{0}
\makeatletter
\newcommand{\rmnum}[1]{\romannumeral #1}
\newcommand{\Rmnum}[1]{\expandafter\@slowromancap\romannumeral #1@}
\makeatother

\newpage

{
\center \bf \large
Supplemental Material for: \\
Observing topological charges and dynamical bulk-surface correspondence with ultracold atoms\vspace*{0.1cm}\\
\vspace*{0.0cm}
}

\vspace{4ex}

This supplemental material provides details
of the experimental realization (Sec.~\Rmnum{1}), data processing (Sec.~\Rmnum{2}), numerical calculations (Sec.~\Rmnum{3}) and theoretical background (Sec.~\Rmnum{4}).

\section{\Rmnum{1}. Experimental Realization~\label{Sec1}}

Here we briefly explain the generation of lattice and Raman potentials (Sec.~\Rmnum{1}A), the derivation of the Bloch Hamiltonian for different settings (Sec.~\Rmnum{1}B),
and the calibration of the two phases $\varphi_{1,2}$ (Sec.~\Rmnum{1}C).
More details can be found in Refs.~\cite{topoCharge,WangBaoZong2018,W.S_realization2DSOC}.

\subsection{\Rmnum{1}A. Raman lattice scheme}\label{Sec1_1}

\begin{figure}
\includegraphics[width=0.5\linewidth]{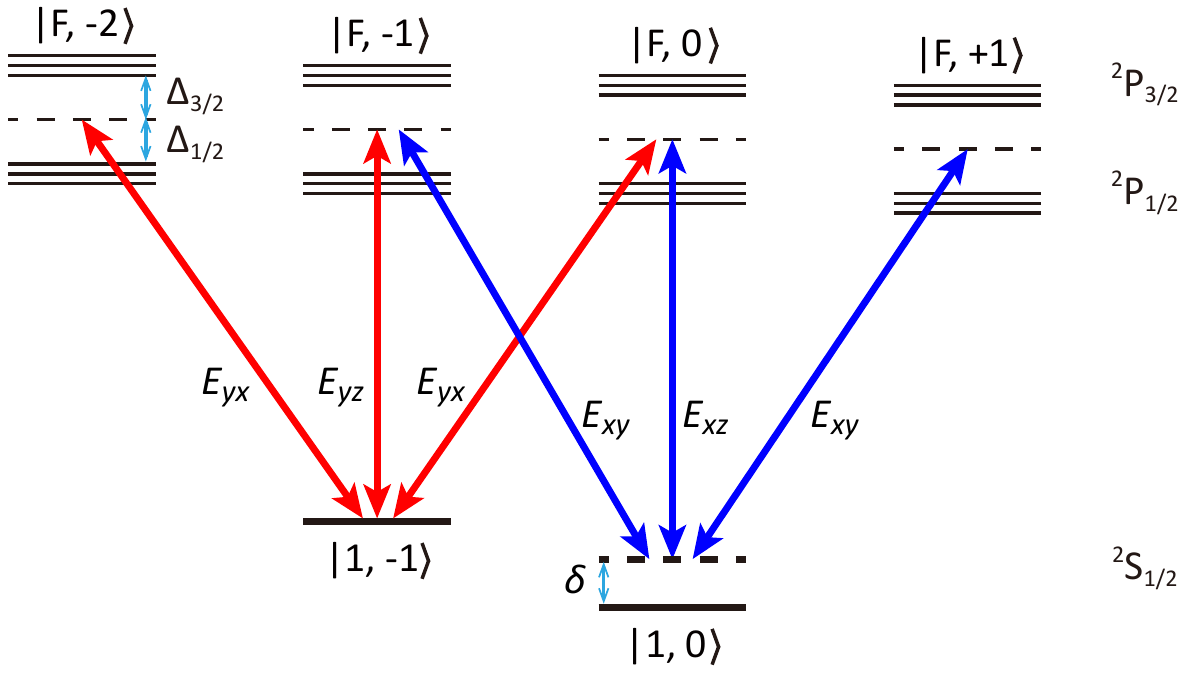}
\caption{Light couplings of both $D_2$ ($5{^{2}S}_{1/2}\to5{^{2}P}_{3/2}$) and $D_1$ ($5^{2}S_{1/2}\to5^{2}P_{1/2}$) transitions for $^{87}$Rb atoms.
}
\label{figS1}
\end{figure}

The experimental setup for $^{87}$Rb Bose gas is shown in Fig.~1a of the main text.
Two magnetic sublevels $|\uparrow\rangle\equiv|F=1,m_F=-1\rangle$ and $|\downarrow\rangle\equiv|1,0\rangle$
are selected by ${10.2} \si{\mega\hertz}$ Zeeman splitting generated via a bias magnetic field of ${14.5}$G applied in $\hat{z}$ direction.
The laser beam $\bm{E}_x$ ($\bm{E}_y$) with wavelength of $\lambda={787} \si{\nano\meter}$ injected from $\hat{x}$ ($\hat{y}$) direction passes through
a high extinction-ratio polarization beam splitting and a $\lambda/2$ wave plates to generate two orthogonally polarized components
$\bm{E}_{xy}$ ($\bm{E}_{yx}$) and $\bm{E}_{xz}$ ($\bm{E}_{yz}$).
A electro-optic modulator (EOM) is placed in front of the mirror $R_x$ ($R_y$) to introduce a phase shift
$\varphi_{1}$ ($\varphi_{2}$) between the two components $\bm{E}_{xy}$ and $\bm{E}_{xz}$ ($\bm{E}_{yx}$ and $\bm{E}_{yz}$).
Hence, the laser beams $\bm{E}_x$ and $\bm{E}_y$ form the standing-wave fields for atoms:
\begin{align}\label{laserField}
{\bm E_x} &=\hat y\left|E_{xy}\right|e^{i(\alpha+\alpha_{L}/2)}\cos(k_0x-\alpha_{L}/2)+\hat z\left|E_{xz}\right|e^{i(\alpha+\alpha_{L}/2+\varphi_{1})}\cos(k_0x-\alpha_{L}/2-\varphi_{1}), \nonumber\\
{\bm E_y}&=\hat x\left|E_{yx}\right|e^{i(\beta+\beta_{L}/2)}\cos(k_0y-\beta_{L}/2)+\hat z\left|E_{yz}\right|e^{i(\beta+\beta_{L}/2+\varphi_{2})}\cos(k_0y-\beta_{L}/2-\varphi_{2}),
\end{align}
where $k_0=2\pi/\lambda$, $\alpha$ and $\beta$ denote the initial phases, and $\alpha_L$ ($\beta_L$) is the phase acquired by ${\bm E}_x$ (${\bm E}_y$) for an additional optical path
from the atom cloud to mirror $R_x$ ($R_y$), then back to the atom cloud.
As shown in Fig.~\ref{figS1},
the lattice and Raman coupling potentials are contributed from both the $D_2$ ($5{^{2}S}_{1/2}\to5{^{2}P}_{3/2}$) and $D_1$ ($5^{2}S_{1/2}\to5^{2}P_{1/2}$) lines.
The total Hamiltonian reads ($\hbar=1$)
\begin{equation}\label{Ham}
H=\left[\frac{{\bm k}^2}{2m}+V_{\rm latt}(x,y)\right]\otimes{\bm 1}+ \Omega_x(x,y)\sigma_x+ \Omega_y(x,y)\sigma_y+m_z\sigma_z,
\end{equation}
where
\begin{eqnarray}
V_{\rm latt}(x,y)=\frac{1}{3}\left(\frac{t_{D_2}^2}{\Delta_{D_2}}+\frac{t_{D_1}^2}{\Delta_{D_1}}\right)(|{\bm E}_{xz}|^2+|{\bm E}_{xy}|^2+|{\bm E}_{yz}|^2+|{\bm E}_{yx}|^2).
\end{eqnarray}
is the square lattice potential, $\Omega_{x,y}$ are the Raman potentials determined by the two Raman couplings
 \begin{align}
\Omega_x=\frac{1}{6}\left(\frac{t_{D_2}^2}{2\Delta_{D_2}}-\frac{t_{D_1}^2}{\Delta_{D_1}}\right)(\hat{e}_z\cdot{\bm E}_{xz})(\hat{e}_-\cdot{\bm E}_{yx}), \quad\quad
\Omega_y=\frac{1}{6}\left(\frac{t_{D_2}^2}{2\Delta_{D_2}}-\frac{t_{D_1}^2}{\Delta_{D_1}}\right)(\hat{e}_+\cdot{\bm E}_{xy})(\hat{e}_z\cdot{\bm E}_{yz}),
\end{align}
and $m_z=\delta/2$ measures the two-photon
detuning of Raman coupling.
Here $\hat{e}_{\pm}=(\hat{e}_{x}\mp i\hat{e}_{y})/\sqrt{2}$, $t_{D_2}\equiv|\langle J=1/2||e{\bm r}||J'=3/2\rangle|$, $t_{D_1}\equiv|\langle J=1/2||e{\bm r}||J'=1/2\rangle|$, and $t_{D_2}\approx\sqrt{2}t_{D_1}\approx4.227ea_0$~\cite{Rb87data}, with $a_0$ being the Bohr radius.
One can tune the phases  $\varphi_{1,2}$ by EOMs to manipulate the Raman and lattice potentials.

\subsection{\Rmnum{1}B. Symmetric and asymmetric settings}\label{Sec1_2}

We consider three cases of the phases: (i) $(\varphi_{1},\varphi_{2})=(\pi/2,\pi/2)$; (ii) $(\varphi_{1},\varphi_{2})=(0,\pi/2)$;
(iii) $(\varphi_{1},\varphi_{2})=(\pi/2,0)$.

(i) When $(\varphi_{1},\varphi_{2})=(\pi/2,\pi/2)$, the light fields read
\begin{align}
{\bm E}_x&=\hat{y}|E_{xy}|e^{i(\alpha+\alpha_L/2)}\cos(k_0 x-\alpha_L/2)+i\hat{z}|E_{xz}|e^{i(\alpha+\alpha_L/2)}\sin(k_0 x-\alpha_L/2),\nonumber\\
{\bm E}_y&=\hat{x}|E_{yx}|e^{i(\beta+\beta_L/2)}\cos (k_0 y-\beta_L/2)+i\hat{z}|E_{yz}|e^{i(\beta+\beta_L/2)}\sin(k_0 y-\beta_L/2),
\end{align}
which leads to the lattice potential
\begin{align}
V_{\rm latt}(x,y)=V_{0x}\cos^2(k_0 x-\alpha_L/2)+V_{0y}\cos^2(k_0 y-\beta_L/2),
\end{align}
with
\begin{equation}\label{V0xy}
V_{0x/0y}=\frac{t_{D_1}^2}{3}\left(\frac{1}{|\Delta_{D_1}|}-\frac{2}{|\Delta_{D_2}|}\right)(|E_{xy/yx}|^2-|E_{xz/yz}|^2),
\end{equation}
and the Raman coupling potentials
\begin{align}
{\Omega}_x=\Omega_{0x}\sin(k_0 x-\alpha_L/2)\cos (k_0 y-\beta_L/2),\quad\quad{\Omega}_y=\Omega_{0y}\cos(k_0 x-\alpha_L/2)\sin(k_0 y-\beta_L/2),
\end{align}
with
\begin{equation}\label{M0xy}
\Omega_{0x/0y}=\frac{t_{D_1}^2}{6\sqrt{2}}\left(\frac{1}{|\Delta_{D_1}|}+\frac{1}{|\Delta_{D_2}|}\right)|E_{xz/xy}||E_{yx/yz}|.
\end{equation}
In the tight-binding limit and only considering $s$-bands, the Hamiltonian (\ref{Ham}) takes the form in the momentum space
$H=\sum_{\bm q} {\Psi}^\dagger_{\bm q}{\cal H}{\Psi}_{\bm q}$~\cite{topoCharge,WangBaoZong2018},
where ${\Psi}_{\bm q}=({c}_{{\bm q}\uparrow},{c}_{{\bm q}\downarrow})$ and
\begin{equation}\label{Hq}
{\cal H}^{({\rm i})}=[m_z-2t_0(\cos q_xa+\cos q_y a)]{\sigma}_z+2t_{\rm so}\sin q_xa{\sigma}_y+2t_{\rm so}\sin q_ya{\sigma}_x
\end{equation}
is exactly the 2D QAH model.
Here  $a$ is the lattice constant with $a=\lambda/2$, ${\bm q}$ is the Bloch momentum,
and  $t_0$ and $t_{\rm so}$ are, respectively, the spin-conserved and spin-flipped hopping coefficients
\begin{align}
t_0&=-\int d{\bm r}\phi_{s}(x,y)\left[\frac{{\bm k}^2}{2m}+V_{\rm latt}({\bm r})\right]\phi_{s}(x-a,y),\nonumber\\
t_{\rm so}&=\Omega_{0}\int d{\bm r}\phi_{s}(x,y)\cos(k_0y)\sin(k_0x)\phi_{s}(x-a,y),
\end{align}
where $\phi_{s}(x,y)$ denotes the Wannier function and  $\Omega_{0x}=\Omega_{0y}=\Omega_{0}$.

(ii) When $(\varphi_{1},\varphi_{2})=(0,\pi/2)$, the light fields become
\begin{align}
{\bm E}_x&=\hat{y}|E_{xy}|e^{i(\alpha+\alpha_L/2)}\cos(k_0 x-\alpha_L/2)+\hat{z}|E_{xz}|e^{i(\alpha+\alpha_L/2)}\cos(k_0 x-\alpha_L/2),\nonumber\\
{\bm E}_y&=\hat{x}|E_{yx}|e^{i(\beta+\beta_L/2)}\cos (k_0 y-\beta_L/2)+i\hat{z}|E_{yz}|e^{i(\beta+\beta_L/2)}\sin(k_0 y-\beta_L/2).
\end{align}
The lattice potential is
\begin{align}
V_{\rm latt}(x,y)=V'_{0x}\cos^2(k_0 x-\alpha_L/2)+V_{0y}\cos^2(k_0 y-\beta_L/2),
\end{align}
with $V_{0y}$ being defined as in Eq.~(\ref{V0xy}) and
$V'_{0x}=\frac{t_{D_1}^2}{3}\left(\frac{1}{|\Delta_{D_1}|}-\frac{2}{|\Delta_{D_2}|}\right)(|E_{xy}|^2+|E_{xz}|^2)$.
The Raman potentials are
 \begin{align}
\Omega_x=0,\quad\Omega_y=\Omega_{0x}\cos(k_0 x-\alpha_L/2)\cos (k_0 y-\beta_L/2)+\Omega_{0y}\cos(k_0 x-\alpha_L/2)\sin(k_0 y-\beta_L/2).
 \end{align}
 which generate spin-flipped hopping only in $\hat{y}$-direction, accompanied with on-site spin flipping.
 In the tight-binding limit, the Bloch Hamiltonian reads~\cite{topoCharge}
\begin{equation}
 {\cal H}^{({\rm ii})}=[m_z-2(t^x_0\cos q_xa+t_0^y\cos q_y a)]\sigma_z+2t^x_{\rm so}\sin q_ya\sigma_x+m_y\sigma_y,
\end{equation}
where
\begin{equation}
m_y=\Omega_{0x}\int d{\bm r}\phi_{s}({\bm r})\cos(k_0x)\cos(k_0y)\phi_{s}({\bm r}).
\end{equation}

(iii) When $(\varphi_{1},\varphi_{2})=(\pi/2,0)$, the light fields become
\begin{align}
{\bm E}_x&=\hat{y}|E_{xy}|e^{i(\alpha+\alpha_L/2)}\cos(k_0 x-\alpha_L/2)+i\hat{z}|E_{xz}|e^{i(\alpha+\alpha_L/2)}\sin(k_0 x-\alpha_L/2),\nonumber\\
{\bm E}_y&=\hat{x}|E_{yx}|e^{i(\beta+\beta_L/2)}\cos (k_0 y-\beta_L/2)+\hat{z}|E_{yz}|e^{i(\beta+\beta_L/2)}\cos(k_0 y-\beta_L/2).
\end{align}
The lattice potential is
\begin{align}
V_{\rm latt}(x,y)=V_{0x}\cos^2(k_0 x-\alpha_L/2)+V'_{0y}\cos^2(k_0 y-\beta_L/2),
\end{align}
with $V_{0x}$ being defined as in Eq.~(\ref{V0xy}) and
$V'_{0y}=\frac{t_{D_1}^2}{3}\left(\frac{1}{|\Delta_{D_1}|}-\frac{2}{|\Delta_{D_2}|}\right)(|E_{yx}|^2+|E_{yz}|^2)$.
The Raman potentials are
 \begin{align}
\Omega_x=\Omega_{0x}\sin(k_0 x-\alpha_L/2)\cos (k_0 y-\beta_L/2)-\Omega_{0y}\cos(k_0 x-\alpha_L/2)\cos(k_0 y-\beta_L/2),\quad\Omega_y=0.
 \end{align}
 which generate spin-flipped hopping only in $\hat{x}$-direction, and also on-site spin flipping.
 In the tight-binding limit, the Bloch Hamiltonian reads~\cite{topoCharge}
\begin{equation}
 {\cal H}^{({\rm iii})}=[m_z-2(t^x_0\cos q_xa+t_0^y\cos q_y a)]\sigma_z+2t^y_{\rm so}\sin q_xa\sigma_y+m_x\sigma_x,
\end{equation}
where
\begin{equation}
m_x=\Omega_{0y}\int d{\bm r}\phi_{s}({\bm r})\cos(k_0x)\cos(k_0y)\phi_{s}({\bm r}).
\end{equation}

\begin{figure}
\includegraphics[width=0.5\linewidth]{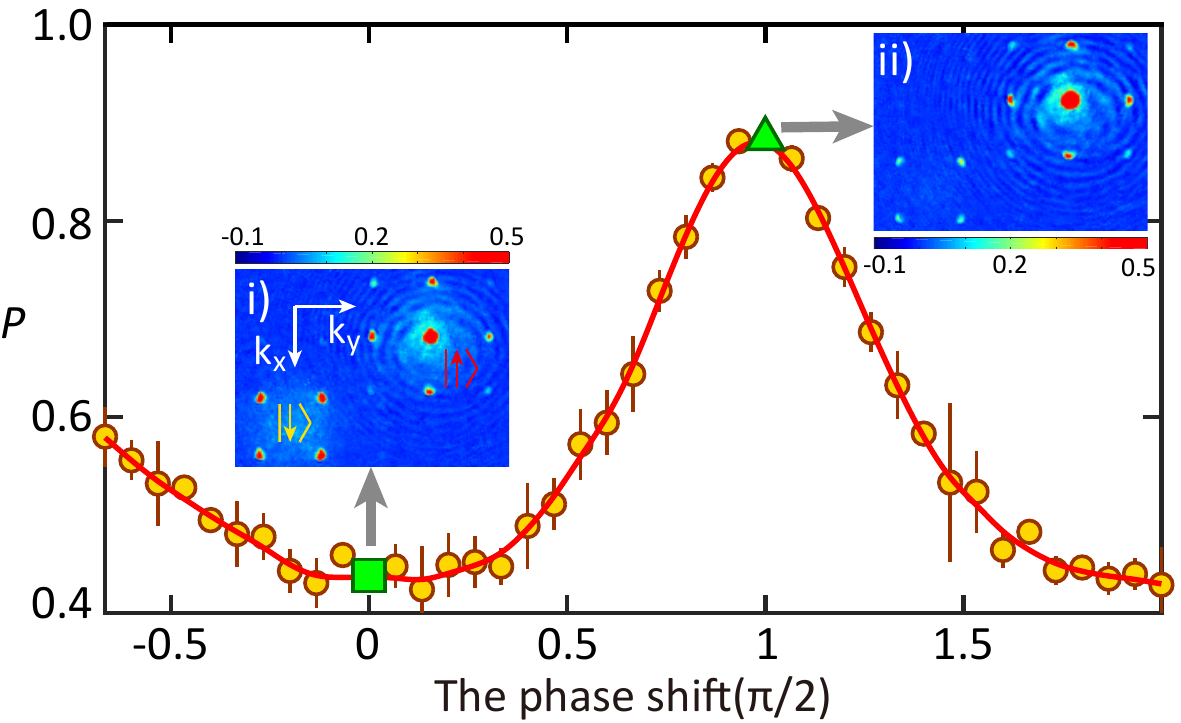}
\caption{Calibration of the phase $\varphi_{1}$.
Spin polarization $P$ of BEC atoms is measured as a function of the phase $\varphi_{1}$ with $\varphi_{2}=\pi/2$.
The circles with error bars are experimental measurements and the red curve is the fitting curve.
The phase $\varphi_{1}$ is tuned to be 0 (or $\pi/2$) when $P$ reaches the minimum (or maximum).
The inset {\romannumeral1}) and {\romannumeral2}) show the spin-resolved TOF imaging of the BEC in the case
corresponding to $\varphi_{1}=0$ (green square) and $\varphi_{1}=\pi/2$ (green triangle), respectively.
The lattice depth $V_{0y}$ is varied from $4E_{\rm r}$ ($\varphi_{1}=\pi/2$) to $5.2E_{\rm r}$ ($\varphi_{1}=0$), while
other parameters are fixed at $V_{0x}=4.0E_{\rm r}$, $\Omega_0=1.0E_{\rm r}$ and $\delta=-0.6E_{\rm r}$.
}
\label{figS2}
\end{figure}

\subsection{\Rmnum{1}C. Calibration of the phases $\varphi_{1,2}$}\label{Sec1_3}

In experiment, we tune the phases $\varphi_{1,2}$ by the voltage applied on EOMs ($\SI{-200}{\volt}\sim\SI{200}{\volt}$).
The instability of $\varphi_{1,2}$, which is limited to approximately $\pi/60$ radians per 10 hours, is controlled by the thermoelectric cooler attached to EOMs.
We calibrate the phases $\varphi_{1,2}$ by measuring the spin polarization of the Bose-Einstein condensate (BEC).

We first prepare about $3\times10^5$ $^{87}$Rb atoms in the Raman lattice with some values of the phases $\varphi_{1,2}$, which are cooled to
condense at the ground state of the lowest band ($q_x=q_y=0$).
We then release the BEC atoms for $\SI{25}{\milli\second}$ and take the spin-resolved TOF imaging to obtain the spin
and momentum distribution of the atomic cloud. We calculate the spin polarization of the BEC $P=\frac{N_{\uparrow}-N_{\downarrow}}
{N_{\uparrow}+N_{\downarrow}}$, where $N_{\uparrow}$ ($N_{\downarrow}$) denotes the total atom number in the spin-up (-down) state,
and accordingly calibrate the values of $\varphi_{1,2}$.
We show an example in Fig.~\ref{figS2}, where the spin polarization $P$ is measured as a function of the phase $\varphi_{1}$ with $\varphi_{2}$ being fixed at $\pi/2$.
The measured variation of $P$ reflects the generation of the transverse magnetization $m_{y}$: when $P$ takes the maximum,
the magnetization $m_{y}\simeq0$, corresponding to $\varphi_{1}=\pi/2$; when $P$ is at the minimum,  $m_{y}$ takes its largest value,
corresponding to $\varphi_{1}=0$.

\section{\Rmnum{2}. Data processing}~\label{Sec2}
Here, we explain the obtainment of the time-averaged spin textures from the time evolution of spin polarization (Sec.~\Rmnum{2}A) and the sign of $h_z$ (Sec.~\Rmnum{2}B).

\subsection{\Rmnum{2}A. Data fitting}~\label{Sec2_1}
\begin{figure}
\includegraphics[width=0.7\linewidth]{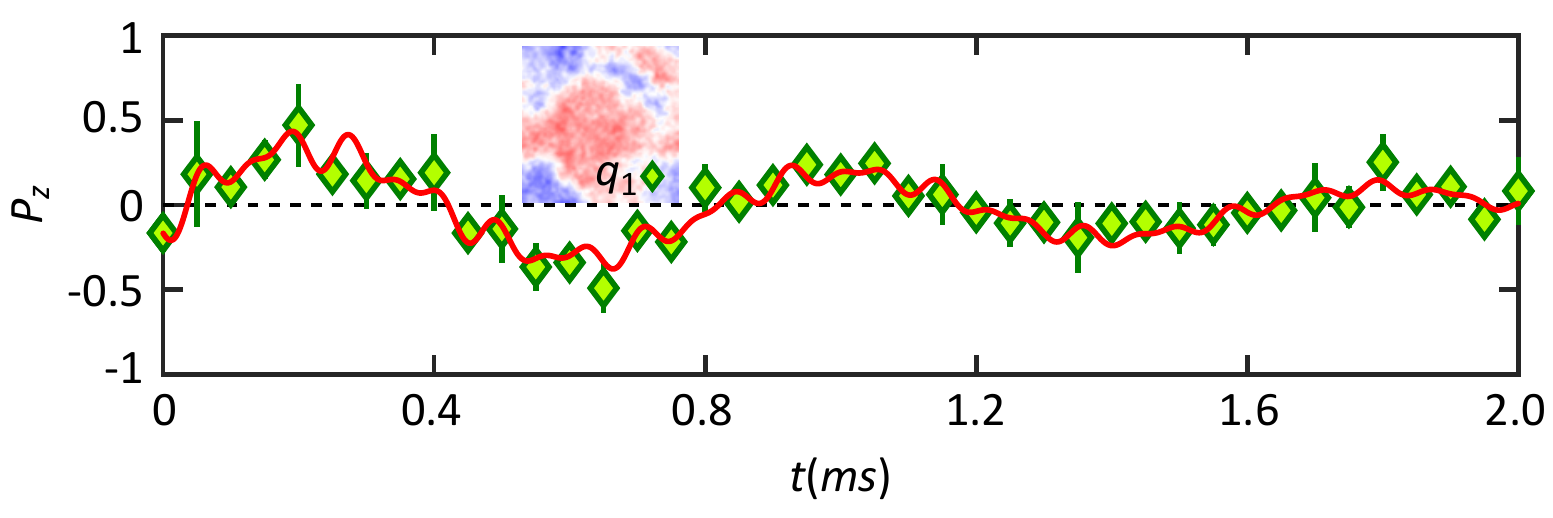}
\caption{Time evolution of spin polarization $\langle\sigma_z(\bm{q},t)\rangle$ at the momentum ${\bm q}_1$ after quenching $h_y$.
The green diamonds with error bars are experimental measurements with the post-quench parameters $V_0=4.0E_{\rm r}$,
$\Omega_0=1.0E_{\rm r}$ and $\delta=-0.2E_{\rm r}$.
The red curve is the fitting curve. The inset is the spin texture at $t=\SI{0.4}{\milli\second}$, where ${\bm q}_1$ is marked.
}
\label{figS3}
\end{figure}

After the measurement,
we fit the time-evolved spin polarization $\langle\sigma_z(\bm{q},t)\rangle$ at each momentum $\bm{q}$ into the combination function
\begin{equation}\label{decay_multi_bandP}
P_{z}(\bm{q},t)=\sum_{i=1}^{3}A_i(\bm{q})\cos(2\pi f_i(\bm{q})t+\phi_i)e^{-\frac{t}{\tau_1}}+B(\bm{q})e^{-\frac{t}{\tau_2}}+D(\bm{q}).
\end{equation}
Here the first $A_i(\bm{q})$-terms denote the damped oscillations with the characteristic damping time $\tau_1$, the second $B(\bm{q})$-term
represents a pure decay with the characteristic time $\tau_2$, and the last $D(\bm{q})$-term is an offset.
Both the decay and damping are internal relaxation induced by the fluctuation of magnetic field and the atom-atom interaction~\cite{uncover_topology,damp_Rabi}.
In the derivation of time-averaged spin textures, we remove those non-ideal effects from Eq.~(\ref{decay_multi_bandP}) and obtain
$\overline{\langle\sigma_{z}(\bm{q})\rangle}_{x,y,z}$ by
\begin{equation}\label{average_multi_bandP}
\overline{\langle \sigma_{z}(\bm{q})\rangle}_{x,y,z}=
\frac{1}{T}\sum_{t=0}^{T} \left(\sum_{i=1}^{3}A_i(\bm{q})\cos(2\pi f_i(\bm{q})t+\phi_i)\right)+B(\bm{q})+D(\bm{q}).
\end{equation}
Here $T$ is taken much longer than the oscillation period. As an example, the measured time-evolved spin polarization
at ${\bm q}_1=(0.66k_0,0.66k_0)$ is shown in Fig.~\ref{figS3} after quenching $h_y$. Experimental data are fitted into
Eq.~(\ref{decay_multi_bandP}) (red curve). According to the fitting parameters, we have $\overline{\langle\sigma_{z}(\bm{q}_1)\rangle}_{x}\approx-0.02$.

The time average of all momentum points in the first BZ are shown in Fig. 3(a) for three quenches. The deviation from numerical results comes from experimental errors induced by mechanical shaking and magnetic field fluctuation, as well as interaction induced decay or dephasing.

\subsection{\Rmnum{2}B. The sign of $h_z$}~\label{Sec2_2}
\begin{figure}
\includegraphics[width=0.7\linewidth]{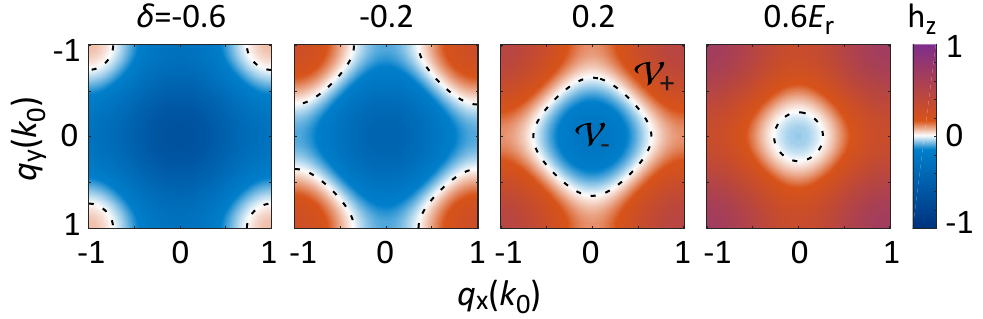}
\caption{$h_z$ for different detuning $\delta$ with $t_0=0.09$. The dash curves are $h_z=0$, which is BIS. The BISs divide the first BZ into two regions: $\mathcal{V_-}$ with $h_z<0$ and $\mathcal{V_+}$ with $h_z>0$.
}
\label{figS4}
\end{figure}

The sign of $h_z=m_z-2t_0(\cos(q_x)+\cos(q_y))$ can be determined by the size of the BISs from the experiments. In the theory [Fig.~\ref{figS4}], the region $\mathcal{V_-}$ as well as the size of the BISs shrinks when the detuning increases from $\delta=-0.6$ to $\delta=0.6$, from which we obtain $h_z<0$ ($h_z>0$) in the region $\mathcal{V_-}$ ($\mathcal{V_+}$).
In the experiment, we measure the time evolution of spin polarization with different detuning and fixed lattice depth and Raman coupling strength in topological nontrivial regime. The measured results are found in Ref.~\cite{uncover_topology}.
The size of the BISs shrinks when the detuning increases, which is consistent with theoretical calculation.
Therefore, $h_z<0$ in the region $\mathcal{V_-}$ and $h_z>0$ in the region $\mathcal{V_+}$.

\section{\Rmnum{3}. Numerical calculations}~\label{Sec3}

We compare our experimental observations to the numerical calculations.
In the calculations,
the post-quench unitary dynamics is described by the time  evolution operator $U(\bm{q},t)=\exp(-iHt)$,
where $H$ is the full-band Hamiltonian (\ref{Ham}) with $\varphi_1=\varphi_2=\pi/2$.
The time-dependent density matrix is then $\rho(\bm{q},t)=U(\bm{q},t)\rho (\bm{q},t=0)U(\bm{q},t)^\dagger$.
Here the initial state reads $\rho(\bm{q},0)=\sum_np_n|n\rangle\langle n|$, where $|n\rangle$ denotes the $n$-th eigenstate of $H$ with pre-quench
parameters and $p_n$ is the population probability at the state $|n\rangle$ determined by the Bose-Einstein statistics $f(E_n;T,\mu)$, with the temperature $T$ being
$150$nk ($100$nk) for quenching $h_{x,y}$ ($h_z$), and the chemical potential $\mu$ being fixed by the particle density.
The time-evolved spin textures are then given by $\langle \sigma_z(\bm{q},t)\rangle={\rm Tr}[\rho(\bm{q},t)\sigma_z]$, and the time averages
$\overline{\langle\sigma_z(\bm{q})\rangle}_{x,y,z}$ are obtained by summing over a long enough time so that they approaches the ones averaged over infinite time.

The time-averaged spin textures in topologically trivial region are also measured, as shown in Fig.\ref{figS4}
with $V_0=4.0E_{\rm r},\Omega_0=1.0E_{\rm r},\delta=-1.0E_{\rm r}$.
One can see that there is no momentum satisfying $\overline{\langle  \sigma_z ({\bm q})\rangle}_{x,y,z}=0$, which characterizes
the trivial phase with Chern number $\mathcal{C}=0$. The observations agree with the numeral calculations with the same experimental parameters (Fig.\ref{figS4}).

\begin{figure}
\includegraphics[width=0.6\linewidth]{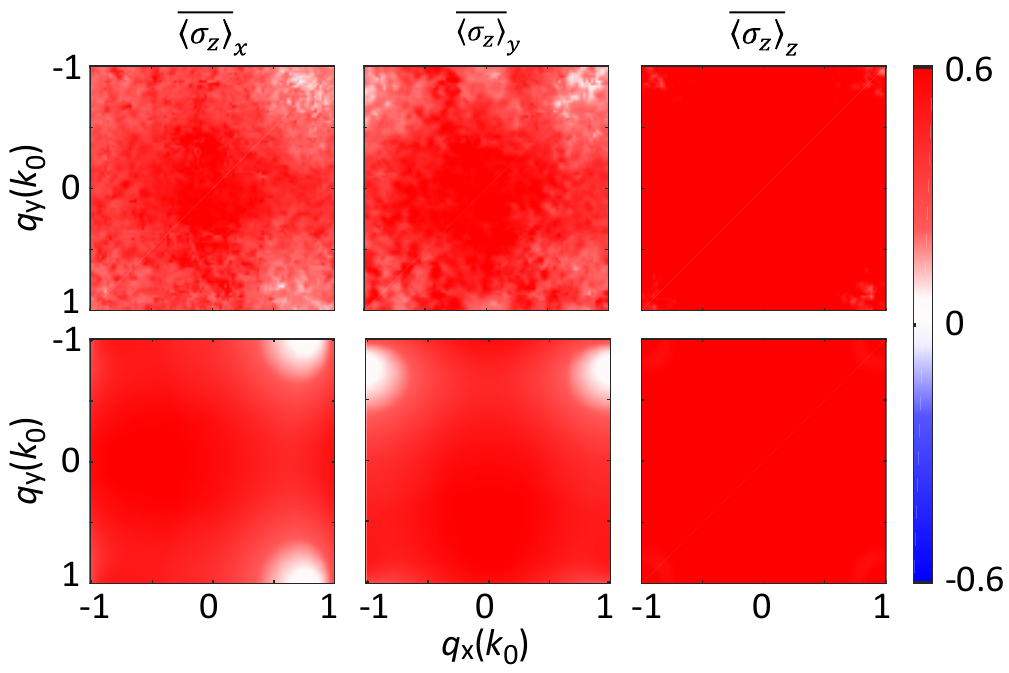}
\caption{Time-averaged spin textures in the topologically trivial phase.
The first (second) row is the experimental measurements (numerical calculations) of $\overline{\langle  \sigma_z ({\bm q})\rangle}$ after quenching $h_{x,y,z}$.
The post-quench parameters are $V_0=4.0E_{\rm r},\Omega_0=1.0E_{\rm r},\delta=-1.0E_{\rm r}$.
}
\label{figS5}
\end{figure}

\section{\Rmnum{4}. Theoretical background}~\label{Sec4}

Here we present briefly the theoretical background of our experiment. Details can be found in Refs.~\cite{windingNumber_zhang,topoCharge,TopoQuenchGeneral}.
In our dynamical classification theory, two essential concepts are introduced: One is band inversion surfaces (BISs) defined as the surfaces with $h_z({\bm k})=0$; the other is topological charges
located at nodes of the spin-orbit (SO) field ${\bm h}_{\rm so}({\bm k})\equiv(h_y,h_x)=0$. We find that the bulk topology is classified  by either the total
charges enclosed by BISs, or the winding of the SO field ${\bm h}_{\rm so}({\bm k})$ on BISs.
We make use of the time-averaged spin textures after quench, which is defined by ($i=x,y,z$)
\begin{align}
\overline{\langle\sigma_{z}({\bm k})\rangle}_i&=\lim_{T\to\infty}\frac{1}{T}\int_{0}^{T} d t\,\mathrm{Tr}\left[\rho_i(0)e^{i\mathcal{H}t}\sigma_{z}e^{-i\mathcal{H}t}\right]
=h_z{\rm Tr}\left[\rho_i(0){\cal H}\right]/E^2,
\end{align}
with $E=\sqrt{h_x^2+h_y^2+h_z^2}$,
to dynamically characterize the BISs, the SO field, the charges, and also the topology.
Here $\rho_i(0)$ denotes the initial state for quenching $h_i$ and $\mathcal{H}$ is the post-quench Bloch Hamiltonian.
In the following, we first give the description of dynamical characterization by deep quenches~\cite{windingNumber_zhang,topoCharge}, and then discuss the shallow quench situation
that our experiment belongs to.

\subsection{A. Deep quenches}
When the quench starts from the extremely deep trivial regime $m_i\to-\infty$, the initial state $\rho_i(0)$ yields ${\rm Tr}\left[\rho_i(0){\cal H}\right]=h_i$ and
the time-averaged spin texture reads
\begin{eqnarray}\label{gamma0i_d}
\overline{\langle\sigma_{z}({\bm k})\rangle}_i=h_{z}({\bm k})h_i({\bm k})/E^{2}({\bm k}).
\end{eqnarray}
One can see that no matter which axis is quenched,
the spin polarization $\overline{\langle \sigma_{z}({\bm k})\rangle }_i$ always vanishes on BISs where $h_{z}({\bm k})=0$.
Hence, a BIS can be dynamically determined by vanishing time-averaged spin polarization independent of the quench axis:
\begin{align}\label{BIS}
\mathrm{BIS}=\{{\bm k}\vert\overline{\langle{\sigma_z}({\bm k})\rangle}_{x,y,z}=0\}.
\end{align}
In particular, $\overline{\langle{\sigma_z}({\bm k})\rangle}_z=0$ occurs only on BISs, so one can find out these surfaces simply by quenching $h_z$.
Besides, the spin texture $\overline{\langle\sigma_{z}({\bm k})\rangle}_{x/y}$ also vanish on the surfaces
with $h_{x/y}({\bm k})=0$ [see Eq.~(\ref{gamma0i_d})]. Accordingly, the nodes of the SO field can be found out by
 $\overline{\langle\sigma_z({\bm k})\rangle }_{x,y}=0$ but $\overline{\langle\sigma_{z}({\bm k})\rangle }_{z}\neq0$.
Furthermore, near a node ${\bm k}={\bm q}_c$,
 the time-averaged spin texture directly reflects the SO field:
$\overline{\langle \sigma_{z}({\bm k})\rangle}_i\big\vert_{{\bm k}\to{\bm q}_c}\simeq h_{i}({\bm k})/h_z({\bm q}_c)$.
Thus, we define a dynamical spin-texture field ${\bm S}({\bm k})$, whose components are
\begin{equation}\label{si_d}
S_i({\bm k})\equiv\frac{{\rm sgn}[h_z({\bm k})]}{{\cal N}_{\bm k}}\overline{\langle\sigma_{z}({\bm k})\rangle}_i,
\end{equation}
{with ${\cal N}_{\bm k}$ being a factor}.
With this result, the topological charge can be dynamically determined by~\cite{topoCharge}
\begin{equation}
{\cal C}={\rm sgn}[J_{{\bm S}}({\bm q}_c)].
\end{equation}
where $J_{{\bm S}}({\bm k})\equiv \det\left[(\partial S_{i}/\partial k_j)\right]$ is Jacobian determinant.

Moreover, we define a dynamical directional derivative field ${\bm g}({\bm k})=(g_y,g_x)$,
whose components are given by
\begin{align}\label{gi}
g_{i}({\bf k})=\frac{1}{{\cal N}_k}\partial_{k_\perp}\overline{\left\langle \sigma_{z}\right\rangle}_{i},
\end{align}
where $k_\perp$ denotes the momentum perpendicular to BIS,
and ${\cal N}_k$ is the normalization factor.
It can be shown that on the BIS ${\bm g}({\bm k})=\hat{\bm h}_{\rm so}({\bm k})$, where
$\hat{\bm h}_{\rm so}({\bm k})\equiv{\bm h}_{\rm so}({\bm k})/|{\bm h}_{\rm so}({\bm k})|$ is the unit SO field.
Thus the topological invariant can be characterized dynamically by
\begin{equation}\label{wd}
{\cal W}=\sum_j\frac{1}{2\pi}\int_{{\rm BIS}_j} \left[g_y({\bm k})dg_x({\bm k})-g_x({\bm k})dg_y({\bm k})\right],
\end{equation}
where the summation is over all the BISs.
The result  manifests itself a highly nontrivial dynamical bulk-surface correspondence~\cite{windingNumber_zhang}.

\subsection{B. Shallow quenches}

When the quench starts from a shallow trivial regime, i.e., $m_i$ is finitely large, we find that the dynamical characterization
is not valid for any shallow quenches~\cite{TopoQuenchGeneral}. Here we give the validity condition for our quench experiment.
For an shallow quench, the time-averaged spin texture read
\begin{eqnarray}\label{gamma0i_s}
\overline{\langle\sigma_{z}({\bm k})\rangle}_i=h_{z}({\bm k}){\rm Tr}\left[\rho_i(0){\cal H}\right]/E^{2}({\bm k}).
\end{eqnarray}
The BISs can still be identified by the vanishing spin polarizations via Eq.~(\ref{BIS}). The directional derivative on BISs becomes
\begin{align}
\partial_{k_\perp}\overline{\left\langle \sigma_{z}\right\rangle}_{i}={\rm Tr}\left[\rho_i(0){\cal H}\right]/E^2
\end{align}
We denote $\widetilde{h}_i\equiv{\rm Tr}\left[\rho_i(0){\cal H}\right]$, and consider the situation that a quench corresponds to adding
an additional constant magnetization $\delta m_i$ in $h_i$ and tuning $\delta m_i$ from $
\delta m_i<-\max\{|h_{i}({\bm k})|\}$ to zero.

To obtain the validity condition, we introduce local transformations $U_i({\bf k})=e^{-i {\bf u}_i({\bf k})}$, with ${\bf u}_i({\bf k})=\sum_{j=x,y,z} u_j^{(i)}\sigma_j$, for each quench,
which rotates the initial state $\rho_{i}$ around the axis $\hat{\bf u}_i\equiv{\bf u}_i/|{\bf u}_i|$ by an angle $2|{\bf u}_i|$
to the fully polarized one $\rho'_{i}=U_i({\bf k})\rho_{i}({\bf k})U^\dagger_i({\bf k})$, with $\sigma_{i}\rho'_{i}=\rho'_{i}$ and
$\sigma_{j\neq i}\rho'_{i}=0$.
We further set $\hat{\bf u}_i$ to be normal to the plane spanned by the $\sigma_i$ axis
and the pre-quench vector field ${\bf h}$ such that $0\leq2|{\bf u}_i|<\pi/2$ and $u^{(i)}_i=0$.
It can be checked that this transformation does not change the topology~\cite{TopoQuenchGeneral}.
Note that after each quench, we return to the same post-quench Hamiltonian ${\cal H}_{\rm post}=\sum_jh_j\sigma_j$.
When quenching $h_i$, we have the pre-quench Hamiltonian ${\cal H}_{\rm pre}^{(i)}={\cal H}_{\rm post}+\delta m_i\sigma_i$.
We notice that the pre-quench vector field should be in the $\sigma_i$ axis after rotation, which gives
\begin{align}\label{Hpre_ri2}
{\cal H}_{\rm pre}^{(i)'}=\left[(h_i+\delta m_i)/\cos2|{\bf u}_i|\right]\sigma_i.
\end{align}
The relation $(h_i+\delta m_i)^2/\cos^22|{\bf u}_i|=(h_i+\delta m_i)^2+\sum_{j\neq i}h_j^2$ leads to
\begin{align}
\delta m_i=-\cot2|{\bf u}_i|\sqrt{E^2-h_i^2}-h_i.
\end{align}
We use the equality
\begin{align}
U_i\sigma_iU_i^\dagger=&\cos2|{\bf u}_i|\sigma_i+i\frac{\sin2|{\bf u}_i|}{|{\bf u}_i|}\sum_{j\neq i}u^{(i)}_j\sigma_i\sigma_j,
\end{align}
and obtain the rotated post-quench Hamiltonian
\begin{align}
{\cal H}_{\rm post}^{(i)'}={\cal H}_{\rm pre}^{(i)'}-\delta m_iU_i\sigma_iU^\dagger_i
=\frac{h_i+\delta m_i\sin^22|{\bf u}_i|}{\cos2|{\bf u}_i|}\sigma_i-i\delta m_i\frac{\sin2|{\bf u}_i|}{|{\bf u}_i|}\sum_{j\neq i}u^{(i)}_j\sigma_i\sigma_j.
\end{align}
Thus, we have ${\rm Tr}[\rho_i {\cal H}_{\rm post}]={\rm Tr}[\rho'_i {\cal H}^{(i)'}_{\rm post}]=\widetilde{h}_i$, where we denote
\begin{align}
\widetilde{h}_i\equiv(h_i+\delta m_i\sin^22|{\bf u}_i|)/\cos2|{\bf u}_i|=h_i\cos2|{\bf u}_i|-\sin2|{\bf u}_i|\sqrt{E^2-h_i^2}.
\end{align}

On the BIS where $h_z({\bf k})=0$, we have
\begin{align}
\widetilde{h}_x=h_x\cos2|{\bf u}_x|-\sin2|{\bf u}_x||h_y|,\quad\quad\widetilde{h}_y=h_y\cos2|{\bf u}_y|-\sin2|{\bf u}_y||h_x|
\end{align}
It can be easily checked that the first terms $\{h_x\cos2|{\bf u}_x|,h_y\cos2|{\bf u}_y|\}$ preserve the topological pattern of ${\bf h}_{\rm so}$ on BISs
while the second terms
$\{-\sin2|{\bf u}_x||h_y|,-\sin2|{\bf u}_y||h_x|\}$ can change it.
Here we assume $\delta m_x=\delta m_y=\delta m$, $h_x=2t_{\rm so}\sin q_x$, $h_y=2t_{\rm so}\sin q_y$, and
$h_z=m_z-2t_0\cos q_x-2t_0\cos q_y$ with $|m_z|<4t_0$. In order to ensure that our dynamical characterization is valid,
we should have $0\leq2|{\bf u}_x|,2|{\bf u}_y|<\pi/4$ and
$\delta m<-2t_{\rm so}(|\sin q_x|+|\sin q_y|)$.
Since $\cos q_x+\cos q_y=m_z/(2t_0)$ on the BIS, we have
\begin{align}\label{mg_condition}
\delta m<\min\left\{-4t_{\rm so}\sqrt{1-\frac{m^2_z}{16t^2_0}},-2t_{\rm so}\right\}.
\end{align}
In our experiment, we have typically $t_{\rm so}=0.5t_0$ and $m_{x,y}=-14t_0$, which satisfies Eq.~(\ref{mg_condition}) very well.

The locations of charges are dynamically determined by $\overline{\langle\sigma_z({\bm k})\rangle }_{x,y}=0$ but $\overline{\langle\sigma_{z}({\bm k})\rangle }_{z}\neq0$,
which gives
\begin{align}
\widetilde{h}_{x,y}({\bm k})=0.
\end{align}
These charges are obviously different from those monopole charges defined by ${\bm h}_{\rm so}({\bm k})=0$, and hence dubbed as {\it dynamical} topological charges~\cite{TopoQuenchGeneral}.
Moreover, the dynamical spin-texture field ${\bm S}({\bf k})$ defined in Eq.~(\ref{si_d}) characterizes these dynamical topological charges.
We have proved that the winding of $(\widetilde{h}_y,\widetilde{h}_x)$ along BISs classifies the bulk topology if the condition (\ref{mg_condition})
is satisfied.
It is then expected that under the same condition, the total dynamical topological charges enclosed by BISs can be used for the dynamical characterization.


\end{document}